\documentclass{jfm}
\usepackage{bm,amsmath,amssymb,natbib,graphicx,mathrsfs}
\usepackage{array,epic,eepic,color,upgreek}

\graphicspath{{./Figs/}}

\newcommand{\W}{Wi}

\newcommand{\R}{Re}
\newcommand{\T}{\mathbb{\uptau}}

\newlength{\piclen}
\title[Dynamics of viscoelastic pipe flow in the maximum drag reduction limit]
{Dynamics of viscoelastic pipe flow in the maximum drag reduction limit}

\author[Jose M. Lopez, George H. Choueiri and Bj\"{o}rn Hof] {J\ls O\ls
  S\ls E\ns M.\ns L\ls O\ls P\ls E\ls Z, G\ls E\ls O\ls R\ls G\ls E\ns H.\ns 
   C\ls H\ls O\ls U\ls E\ls I\ls R\ls I, B\ls J\ls \"{O}\ls R\ls N\ns H\ls O\ls F}

\affiliation{ Institute of Science and Technology Austria, 3400 Klosterneuburg, Austria\\[\affilskip] }

\pubyear{} \volume{} \pagerange{}
\date{\today}
\begin{document}

\maketitle

\begin{abstract} 

  Polymer additives can substantially reduce the drag of turbulent flows and the upper limit, the so called ``maximum drag reduction'' (MDR) asymptote is
  universal, i.e. independent of the type of polymer and solvent used.
  Until recently, the consensus was that, in this limit, flows are in a marginal state where only a minimal level of turbulence activity persists.
  Observations in direct numerical simulations using minimal sized channels appeared to support this view and reported long ``hibernation'' periods where turbulence is
  marginalized. In simulations of pipe flow we find that, indeed, with increasing Weissenberg number (Wi), turbulence expresses long periods of hibernation
  if the domain size is small. However, with increasing pipe length, the temporal hibernation continuously alters to spatio-temporal intermittency and here
  the flow consists of turbulent puffs surrounded by laminar flow. Moreover, upon an increase in Wi, the flow fully relaminarises, in agreement with recent experiments.
  At even larger Wi, a different instability is encountered  causing a drag increase towards MDR. Our findings hence link earlier minimal flow unit simulations with
  recent experiments and confirm that the addition of polymers initially suppresses Newtonian turbulence and leads to a reverse transition. The MDR state on the other hand
  results from a separate instability and the underlying dynamics corresponds to the recently proposed state of elasto-inertial-turbulence (EIT). 

\end{abstract}

%%%%%%%%%%%%%%%%%%%%%%%%%%%%%%%%%%%%%%%%%%%%%%%%%%%%%%%%%%%%%%%%%%%%%%
\section{Introduction}

The addition of small amounts of polymers to a turbulent flow is
known to be one of the most efficient drag reduction technologies. Since its
discovery by~\cite{Toms48}, it has been extensively used to mitigate friction losses
in the pipeline transportation of turbulent fluids.
Polymer drag reduction has also become the subject
of widespread research aimed at understanding the physics underlying
this phenomenon~\citep[see e.g. review by][]{whMu08}. The amount of drag reduction
that is achieved increases with increasing polymer concentration,
but it eventually saturates at an upper limit known as the maximum
drag reduction (MDR) or Virk's asymptote. A remarkable feature of
this asympotic limit is its universality, i.e. it is independent of
polymer type and properties. While first reports on MDR trace
back to the seventies~\citep{Virk70}, a consensus about the nature of
this universality is still lacking. The usual observation of a continuous decrease
in the friction factor with increasing polymer concentration and the eventual saturation to
MDR has led to the interpretation of MDR as a marginal state of turbulence.
However, why turbulence persists and does not fully relaminarise, even though polymers obviously have the
tendency to subdue turbulence, has remained an open question.

The interpretation of MDR as a marginal turbulent state
has recently found support in direct numerical simulations using the FENE-P
(finitely extensible nonlinear elastic-Peterlin) model to describe the polymers
dynamics.~\cite{XiGra10a,XiGra10b,XiGra12a,XiGra12b},
henceforth referred to as X\&G,  performed simulations in a minimal channel
and observed that viscoelastic turbulence
is characterized by the alternation between intervals of high and low friction.
The latter intervals, which they called hibernating turbulence,
were found to share several structural
and statistical features with MDR. Since the frequency and duration of these
intervals increased gradually with increasing polymer elasticity,
they proposed that MDR might be a marginal state of hibernating turbulence  
whose energy cannot be further reduced by polymer activity.
An alternative explanation to the MDR phenomenon was given by~\cite{Samanta13}.
By combining experiments in pipe flow and simulations in channel
flow, they reported the existence of a secondary instability driven by the interplay
between elasticity and inertia at high polymer concentration. Such instability,
which was called elasto-inertial instability (EII),  sets in at Reynolds numbers
below those at which the transition to turbulence occurs in Newtonian flows,
providing an explanation to the early turbulence phenomenon often
observed in experiments. In addition, the experiments showed that
the friction factor associated with the state resulting from the EII,
named elasto-inertial turbulence (EIT), agrees well with that of the
Virk's asymptote. On this basis, the authors suggested that turbulent drag
reduction is eventually limited by the EII,  which prevents
flows from relaminarising, and that the observed MDR friction factor values
are simply the natural drag levels of EIT.

To test these theories,~\cite{Cho18}, hereafter C,L\&H,
investigated the effect of increasing the polymer concentration
on turbulent pipe flow in experiments at
constant Reynolds numbers. Surprisingly, for not too large Reynolds numbers,
the addition of polymers resulted in full relaminarisation.
Here, shear rates and concentrations were moderate,
so that the EII had not occurred yet while Newtonian turbulence was fully suppressed.
Further addition of polymers, however, destabilised the laminar flow and triggered the EII.
Subsequently, the drag increased and the MDR asymptote was approached from the
laminar limit. This scenario strongly suggests that MDR is a state disconnected from Newtonian turbulence,
thereby supporting the theory that MDR is caused by the EII. On the other
hand, the authors observed that prior to relaminarisation the flow
becomes spatio temporally intermittent and consists of slugs and puffs.
This is in principle in line with the temporal intermittency observed by
X\&G. The main difference is that they proposed that the low
drag (or hibernating) phases correspond to the eventual MDR state,
whereas the intermittency in time and space observed by  C,L\&H
is part of a reverse transition and not the asymptotic state. To clarify this point,
we carry out direct numerical simulations of viscoelastic pipe flow, using
short streamwise domain length (twice as long as in~X\&G), and following a path
in parameter space comparable to that of  C,L\&H. As will be shown below,
the dynamical scenario is in good agreement with that of X\&G in that  
low drag periods become longer and longer and appear to approach some asymptotic level
as the Weissenberg number ($\W$) increases. However, for even
larger $\W$, the flow abruptly relaminarises.

Moreover, when the small computational domain is increased to more
realistic sizes, i.e. pipe lengths sufficiently large to contain a puff, 
the temporal intermittency changes to spatio-temporal intermittency,
revealing that, as reported in the experiments by  C,L\&H,
indeed, a reverse transition occurs with increasing $\W$. At the same time,
the approach towards an almost constant drag level reported by
X\&G, and also found in the small domains in the present study, does
not persist in the large domains. Instead, the flow returns to
intermittent puffs and subsequently fully relaminarises. For even larger $\W$,
an instability occurs that, like in the experiments,
leads to a separate fluctuating dynamical state. Our computations hence
qualitatively agree with the experiments of C,L\&H. While the dominant
flow structures reported in experiments of EIT are large scale
streamwise streaks, in simulations of EIT~\citep{Samanta13,dubief13} only
small near wall spanwise oriented vortical structures were found. In the present
case we find the same near wall spanwise vortical structures. These
structures are found to be localised and they give rise to large scale
streamwise streaks, similar to those observed in experiments.

\section{Problem formulation and numerical methods}\label{sec:Problem}

We investigate numerically the dynamics of a dilute polymer solution
flowing through a straight circular pipe at a constant flow rate.
Polymer dynamics is modeled using the FENE-P model~\citep{Bird80}.
Individual polymer molecules are represented
in this model as two inertialess spherical beads connected by a straight
non-linear spring. The orientation and elongation of each
polymer molecule is determined by the end-to-end vector
$\mathbf{q}$ connecting the two beads.
The ensemble average of the tensorial product of all end-to-end
vectors defines a positive-definite symmetric polymer conformation tensor,
$\mathbf{C}_{ij} = <\mathbf{q}_i \otimes \mathbf{q}_j>$, which allows the problem
to be formulated from a continuum medium approach. 

\subsection{Governing equations and dimensionless parameters}\label{sec:Eq}

The governing equations are presented directly in dimensionless form. The pipe
radius $R$, the laminar centreline velocity $u_{lc}$ and the dynamic pressure $\rho u_{lc}^2$
were chosen as characteristic scales for length, velocity and  pressure respectively.
$\mathbf{q}$ was normalized with $\sqrt{kT_e/H}$, where $k$ denotes the
Boltzmann constant, $T_e$ is the absolute temperature and $H$ is the spring constant.
The maximum polymer extension is indicated by the dimensionless parameter $L=q_0/\sqrt{kT_e/H}$,
where $q_0$ is the maximum separation between beads allowed by the spring.  
Cylindrical coordinates $(z,\theta,r)$ are used.

\noindent The temporal evolution of $\mathbf{C}_{ij}$ is obtained by solving the following constitutive equation
\begin{equation}\label{const_eq}
  \begin{split}
    \partial_t\mathbf{C}_{ij}+\mathbf{v}\cdot\nabla\mathbf{C}_{ij} =
    \mathbf{C}_{ij}\cdot\nabla\mathbf{v} + (\nabla\mathbf{v})^T\cdot\mathbf{C}_{ij}-\T_{ij},\\
    i = z,\theta,r \qquad j = z,\theta,r,
  \end{split}
\end{equation}
where $\mathbf{v}=(u,v,w) $ is the velocity vector field and $\T_{ij}$ is the polymer stress tensor.
The first two terms on the right hand side of equation \eqref{const_eq} model polymer stretching due to hydrodynamic forces,
whereas $\T_{ij}$ represents the relaxation forces bringing the polymers back to its equilibrium configuration. 
$\T_{ij}$ is computed using the Peterlin closure
\begin{equation}\label{Peterlin}
  \T_{ij} = \frac{1}{\W}(\frac{\mathbf{C}_{ij}}{1-\frac{tr(\mathbf{C}_{ij})}{L^2}}-\mathbf{I}),
\end{equation}
where $tr(\mathbf{C}_{ij})$ denotes the trace of the polymer conformation tensor,
$\mathbf{I}$ is the unit tensor and $\W$ is the Weissenberg number; a dimensionless number
quantifying the ratio of the polymer relaxation time $\lambda$ to the characteristic flow time scale  $R/u_{lc}$.    

\noindent The fluid motion is governed by the continuity and Navier-Stokes equations 
%\begin{equation}\label{NSeq}
\begin{align}
  \nabla \cdot \mathbf{v} = 0,\label{continuity_eq}\\
  \partial_t\mathbf{\mathbf{v}} + \mathbf{v}\cdot\nabla\mathbf{v} =
  -\nabla P + \frac{\beta}{\R} \nabla^2\mathbf{v} + \frac{(1-\beta)}{\R} \nabla\cdot\T_{ij},\label{NS_eq}
  %\end{equation}
\end{align}
where $P$ is the pressure, $\beta = \nu_s/\nu$ measures the relative importance between the solvent viscosity $\nu_s$ and
the viscosity of the solution at zero shear rate $\nu$, and $\R = u_{lc} R/\nu$ is the Reynolds number.
Polymers modify the dynamics of Newtonian flows through polymer stresses. These are incorporated
into the conventional Navier-Stokes equation through the divergence of the polymer
stress tensor. The $(1-\beta)$ prefactor multiplying this term indicates 
the contribution of the polymers to the total viscosity and must be small for a
dilute polymer solution. Periodic boundary conditions are used in the streamwise $z$
and azimuthal $\theta$ directions, whereas the no-slip condition is imposed at the pipe wall $r=R$.

In all simulations presented in this paper the Reynolds number
was fixed to $\R=3500$, for which the flow is turbulent
in the Newtonian case, and $\W$ was used as  control parameter.
We also fixed $\beta$ to $0.9$ which is the value corresponding to the experiments
of~C,L\&H at a concentration of $90$ ppm. Given the values of $\beta$ and $L$, polymers
can be characterized by their extensibility number $E_x=2L^2(1-\beta)/3\beta$~\citep{XiGra10b}.
For our simulations we have considered two different polymers with very
different extensibilities. The maximum extension of the first polymer type, $L=30$, was chosen so that
its extensibility number, $E_x=66.6$, coincides with one of the cases presented in~\cite{XiGra10b}.
The second polymer type has a very high extensibility, $E_x=2962.96$ for $L=200$, and it corresponds
to the parameters used in simulations of elasto-inertial turbulence by~\cite{dubief13}.
These two cases will be henceforth referred to as moderate extensibility ME and
large extensibility LE cases respectively.

\subsection{Numerical methods}\label{sec:NM}

The governing equations are solved in primitive variables using a highly scalable
pseudo-spectral solver recently developed in-house by our research group.  The code is parallelized
using a combination of the MPI and OpenMP programming models~\citep[see][for further details]{ShiRaHoAv15}.
Spatial discretization in the two periodic directions, $z$ and $\theta$,
is accomplished via Fourier-Galerkin expansions, whereas central
finite differences on a Gauss-Lobatto-Chebyshev grid are used in $r$.
Pressure and velocity in equation \eqref{NS_eq} are decoupled through
a Pressure Poisson Equation (PPE) formulation. An influence matrix is used
to impose the free divergence boundary condition directly on velocity, thereby
avoiding the use of artificial pressure boundary conditions. 
The equations for the azimuthal and radial velocity components $v$ and $w$
are decoupled using the change of variables, $u_+=w+iv$
and $u_-=w-iv$~\citep{OrPa83}.

\noindent The time integration was carried out using a second order accurate
predictor-corrector scheme based on the Crank-Nicolson method~\citep{openpipeflow}. 
For a generic variable $\mathbf{X}$ at a time $n$ the predictor equation reads
\begin{equation}\label{predictor}
(\frac{1}{\delta t}-ic\nabla^2) \mathbf{X}_1^{n+1} = (\frac{1}{\delta t}+(1-ic)\nabla^2) \mathbf{X}^n+\mathbf{N}^n,
\end{equation}
where $\mathbf{N}$ denotes the non-linear terms,  $\delta t$ is the time step size and
the constant $ic$ defines the implicitness of the method ($ic=0.5$ in our simulations).
The initial estimate $\mathbf{X}_1^{n+1}$ is then refined following an iterative correction procedure. At each
corrector iteration the non-linear terms are re-evaluated and $\mathbf{X}_j^{n+1}$ is obtained
solving the following equation
\begin{equation}\label{corrector}
  (\frac{1}{\delta t}-ic\nabla^2) \mathbf{X}_{k+1}^{n+1} = (\frac{1}{\delta t}+(1-ic)\nabla^2) \mathbf{X}^n+ic\mathbf{N}_k^{n+1}+(1-ic)\mathbf{N}^n,
\end{equation}
where $k=1,2,...$ The iteration loop stops when $||\mathbf{X}_{k+1}^{n+1} - \mathbf{X}_{k}^{n+1}|| \leq 10^{-6}$. Convergence usually
occurs after one iteration of the corrector step. The additional computational cost of computing the advective terms twice at each
time step is compensated by the larger $\delta t$ allowed by this temporal scheme in comparison with other conventional methods.
The source terms in equation \eqref{const_eq} and the term containing the divergence of the polymer stress tensor in equation \eqref{NS_eq} are
treated as non-linear terms. Note that equation \eqref{const_eq} is hyperbolic and does not have any diffusive term ($\nabla^2 \mathbf{X}$).
This lack of dissipation leads to numerical error accumulation which often causes spourious instabilities and numerical breakdown.
To avoid these problems we incorporate a small amount of artificial diffusion to our simulations which enhances numerical
stability. This is accomplished by adding a laplacian term $\frac{1}{\R S_c} \nabla^2 \mathbf{C}_{ij}$ to
the right hand side of equation \eqref{const_eq}, where $S_c=\nu/\kappa$ is the Schmidt number quantifying the
ratio between the viscous and artificial diffusivities. In all simulations presented in this paper
the Schmidt number is fixed to $S_c=0.5$. This yields an artificial diffusion coefficient
$\frac{1}{Re S_c} \sim O(10^{-4})$ which is of same order of magnitude as in~\cite{XiGra10b},
and quite below those of early works, e.g.~\cite{ptasinski03,Suresh97}, where $\frac{1}{Re S_c} \sim O(10^{-2})$.
With the inclusion of this laplacian term two boundary conditions are needed:
as suggested in~\cite{Beris99} we impose that
$\mathbf{C}_{ij}$ at $r=R$ must be the same as without artificial diffusion, whereas symmetry
boundary conditions are used at $r=0$.

\noindent The numerical resolution of the simulations presented in this paper
is shown in table~\ref{res}. $\delta t$ is dynamically adjusted
to ensure that the Courant-Friedrichs-Lewy (CFL) condition always remains below $0.25$.
\begin{table}
  \begin{center}
    \begin{tabular}{c c c c c}
      Section & pipe length (R) & $m_r$ & $m_{\theta}$ & $m_z$ \\\hline
      \S4  & 10  & 64 & 64 & 128 \\
      \S5  & 20  & 64 & 64 & 256 \\
      \S5  & 40  & 64 & 64 & 512 \\
      \S5  & 100 & 64 & 64 & 1280\\
      \S6 (EIT)  & 10 & 64 & 100 & 256\\
      \S6 (EIT)  & 40 & 64 & 100 & 1280\\\hline
    \end{tabular}
  \end{center}
  \caption{Number of radial nodes, $m_r$, and Fourier modes, $m_{\theta}$ and $m_z$, used in the simulations.}
  \label{res}
\end{table}

\section{Dynamics of viscoelastic pipe flow turbulence in short computational domains}\label{sec:short}

\begin{figure}\setlength{\piclen}{\linewidth}
  \begin{center}
    \begin{tabular}{c}
      $(a)$ \\
      \includegraphics[width=\piclen]{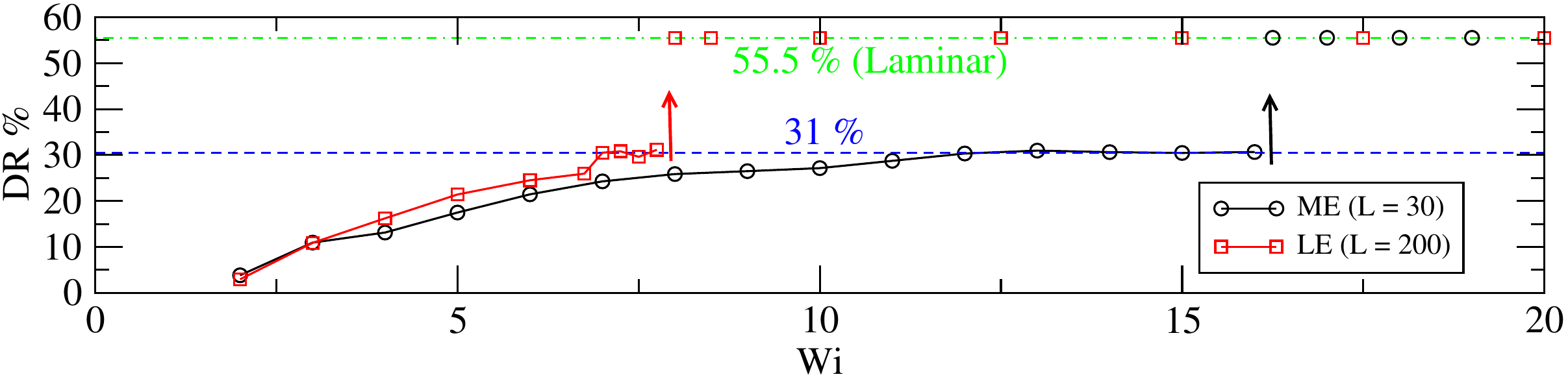} \\
      $(b)$ \\
      \includegraphics[width=\piclen]{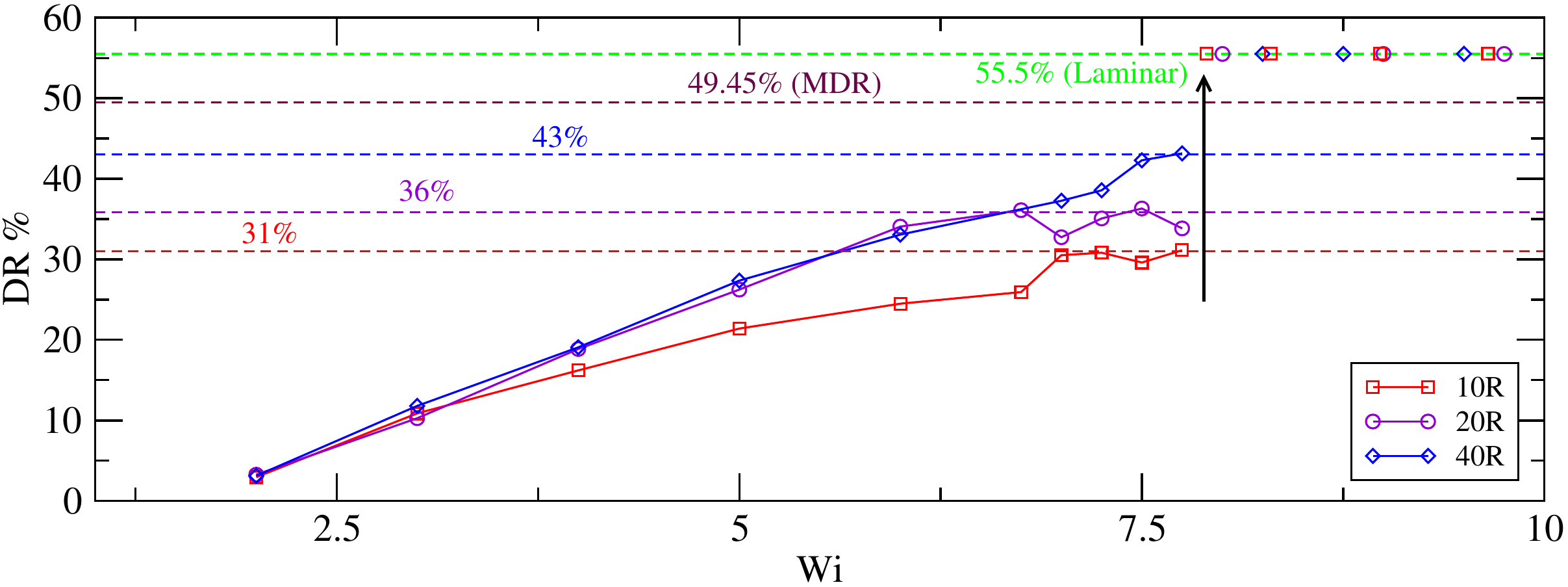}\\
    \end{tabular}
  \end{center}
  \caption{ (Color online) Evolution of the drag reduction percentage $DR\%$ with increasing $\W$ up to relaminarisation
    occurs in pipe flow simulations
    performed at $Re=3500$. $(a)$ Simulations carried out in a $10R$ long pipe using two polymers
    with different extensibilities: LE (large extensibility, $L=200$) and ME (moderate extensibility, $L=30$) .
    $(b)$ Variation of $DR\%$ as the pipe length is
    varied for the LE case.} 
  \label{fig:drag_reduction_simu}
\end{figure}

Because of the additional equations for $C_{ij}$ and $\tau_{ij}$,
viscoelastic turbulence simulations are in computational terms
far more demanding than Newtonian simulations. A common approach to minimise
the computational cost is to choose the smallest
domain size that computes reasonably accurate
dynamics. On that basis, we have set the pipe length to $L_z=10R$, which is nearly the minimum size needed in Newtonian pipe
flow simulations to ensure that these are unaffected by streamwise periodicity.
The simulations were performed according to the following procedure. Starting from
a fully turbulent Newtonian solution, we increased
$\W$ progressively by one unit, with the exception of the range $6 \le \W \le 8$ in the LE case,
where  $\W$ was varied in intervals of $0.25$. The simulations were run over
$2000 R/u_{lc}$ time units and as initial condition we used a previously computed
solution with $\W$ close to that being computed. The averaged drag
reduction percentage was calculated as $DR\% = \frac{f_N - f}{f_N}$, where $f_N$ and $f$ are
the friction coefficients for the Newtonian and viscoelastic cases respectively.
The former is given by the Blasius friction law, $f_N = 0.079 \R^{-0.25}$,
whereas the latter is calculated from the Fanning friction formula, $f = \frac{\tau_w}{2 \rho U_b^2}$,
where $U_b$, $\tau_w$ and $\rho$ are the bulk velocity,
average wall shear stress and fluid density respectively. For each $\W$,
a set of $10$ simulations was performed and the drag reduction level was computed by
averaging over the ensemble of the simulations.

As shown in figure~\ref{fig:drag_reduction_simu} $(a)$,
even for two simulations mimicking different polymers, the same qualitative scenario
in terms of drag reduction is obtained. The amount of drag reduction increases
continuously with increasing $\W$ up to a critical threshold 
after which the flow relaminarises. A clear effect of increasing
the maximum polymer extension $L$ is that the dynamics are accelerated:
the polymer with higher extensibility LE produces for the same $\W$ significantly larger
drag reduction than the ME polymer, and it eventually causes
relaminarisation at a much lower value of $\W$,  $\W_{lam} = 7.75$,
than in the ME case, $\W_{lam} = 16$. We note here that, as $\W_{lam}$ is approached,
the simulations become sensitive to the initial condition and turbulence does not
always survive over the time threshold chosen. The critical values for relaminarisation
$\W_{lam}$ given above correspond to the highest values of $\W$ for which turbulence
survives in more than $50\%$ of the simulations performed. There are also
certain ranges of $\W$ at which polymer extensibility does not appear to
play any role. For example, at very low $\W$ ($W \leq 3$),
the degree of polymer stretching is low and both polymers, despite having
very different extensibility, produce nearly the same drag reduction.
A much more surprising effect occurs at larger $\W$ prior to relaminarisation.
Here, the drag reduction approaches an almost constant level, $31\%$,
regardless of the polymer extensibility. This levelling off was observed in
the earlier study of~X\&G and suggested as an asymptotic regime (AR).
In the present study, the AR occurs over a narrow range of $\W$,
and since dynamical changes take place faster for higher extensibility,
it is much more evident in the ME case, $12 \le \W  \le 16$,  than in the LE case,
$6.75 \le \W \le 7.75$.  

\begin{figure}\setlength{\piclen}{\linewidth}
  \includegraphics[width=\piclen]{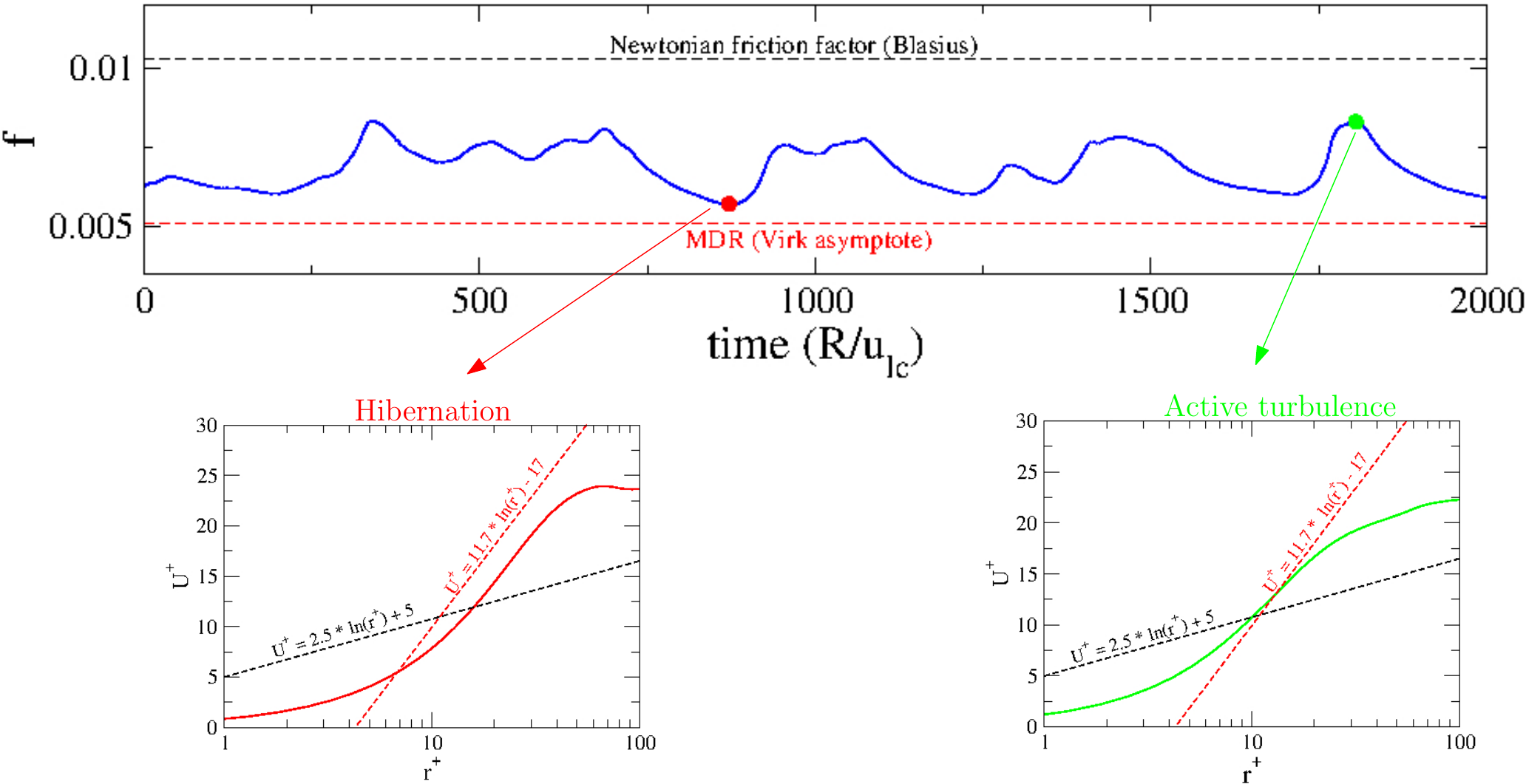} 
  \caption{ (Color online) Temporal intermittency. The top panel shows the temporal
    evolution of the friction factor $f$ for the ME case and
    $\W=13$, whereas the bottom panels illustrate instantaneous mean velocity profiles at low and
    high friction events. Note that here velocity ($U^+$) and radius ($r^+$) are expressed
    in inner units, i.e. normalized with the friction velocity ($u_{\tau}=\sqrt{\tau_w/\rho}$) and
    the viscous length ($\delta_{\nu}=\nu/u_{\tau}$) respectively.}\label{fig:active_hiber}
\end{figure}

A key feature of the dynamics in these simulations is the presence of temporal intermittency,  
with periods of low friction which are interspersed with other periods of higher friction,
as shown in figure~\ref{fig:active_hiber}. These intermittent
dynamics are also in agreement with the simulations of X\&G, who
dubbed the low and high friction intervals as hibernating and active turbulence,
respectively. The frequency and duration of hibernating events increases progressively
with increasing $\W$, and the friction associated with active turbulent events decreases
as $\W$ increases, leading to the gradual growth in average drag reduction shown in the
figure~\ref{fig:drag_reduction_simu} $(a)$. To further illustrate the distinction between
hibernating and active turbulence, the bottom panel in figure~\ref{fig:active_hiber}
shows instantaneous velocity profiles in inner units corresponding to each state. 
The black and red dashed lines in these figures show the universal logarithmic laws
that characterize the mean velocity profile in the logarithmic layer
($30 \lessapprox r^+ \lessapprox 60$, for $\R = 3500$) for wall bounded Newtonian turbulence (Prandtl-K\'arm\'an law) and
viscoelastic turbulence at MDR (Virk's asymptote), respectively. Hibernating events are
characterized by velocity profiles that notably deviate from the Prandtl-K\'arm\'an law and
become nearly parallel to the Virk's asymptote profile throughout the logarithmic layer. By contrast,
in active turbulence events, although friction may be substantially lower than that for pure
Newtonian turbulence, the profile in the log layer has a comparable slope to the Prandtl-K\'arm\'an law.
On the basis of similar observations, it has been argued that states of active turbulence have similar
properties to Newtonian turbulence, whereas hibernating events could be directly connected
to MDR. More specifically, it was suggested that MDR might be a state fully dominated
by hibernation, which is achieved asymptotically as $\W$ is increased~\citep{XiGra10a}.
However, in our simulations, as well as in previous simulations reporting
this intermittent scenario, the flow eventually relaminarises with increasing $\W$ and 
an asymptotic state (the AR) is reached only over a narrow range of $\W$ prior
to relaminarisation. Since the AR  exhibits some  features
of MDR: saturation of the drag reduction level with increasing $\W$
and comparable results are obtained for different polymer properties, it has been
interpreted as the first numerical evidence of MDR. 
However, there is also  evidence which appears to indicate
that the AR does not correspond to MDR. Firstly, while hibernation is prominent
in this regime, active turbulence events also occur frequently,
and so the average drag reduction level at AR ($31\%$) is considerably less than
that of MDR at $Re=3500$ ($49.5\%$). Another distinctive feature is
that in the AR the saturation of drag reduction occurs
over a finite range of $\W$ and upon further increase
in $\W$ the flow relaminarises. In contrast, MDR is a persistent
state and the drag reduction level remains nearly unchanged
as $\W$ increases. Finally, it should also be noted that
temporal intermittent dynamics such as those previously described
have not been reported in experiments at MDR. It is therefore unclear whether
the dynamics of the AR may be related to MDR.

\section{Simulations in larger computational domains: reverse transition}\label{sec:large}
 
\begin{figure}\setlength{\piclen}{0.5\linewidth}
  \begin{center}
    \begin{tabular}{cc}
      $(a)$ & $(b)$\\
      \includegraphics[width=\piclen,height=3cm]{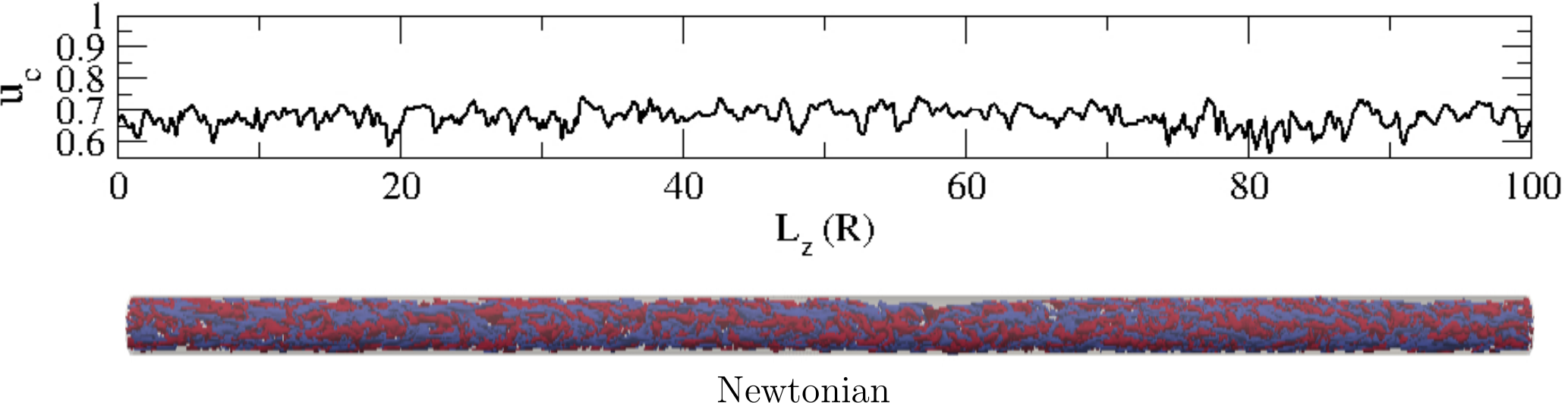} &
      \includegraphics[width=\piclen,height=3cm]{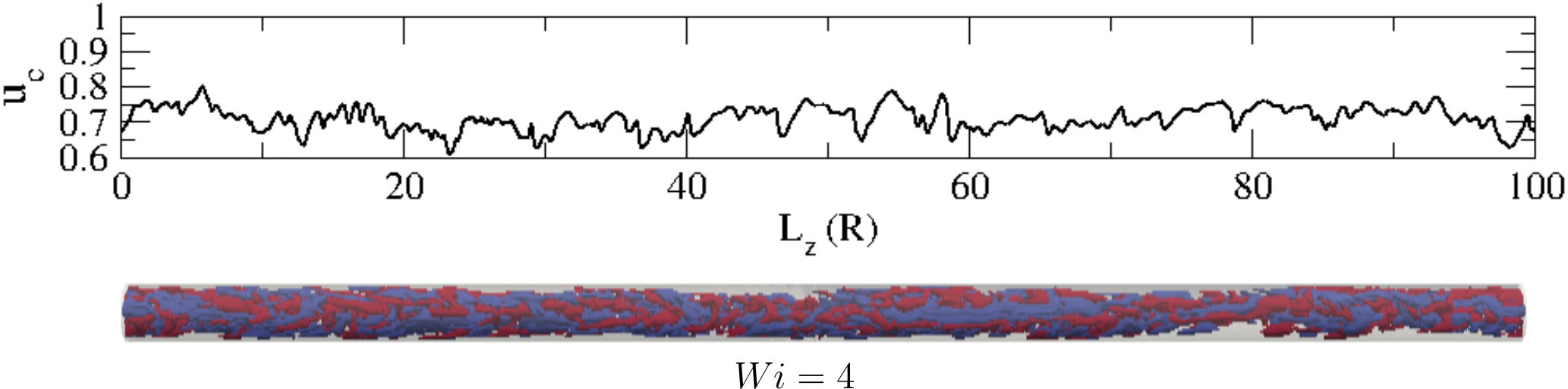}\\
      $(c)$ & $(d)$\\
      \includegraphics[width=\piclen,height=3cm]{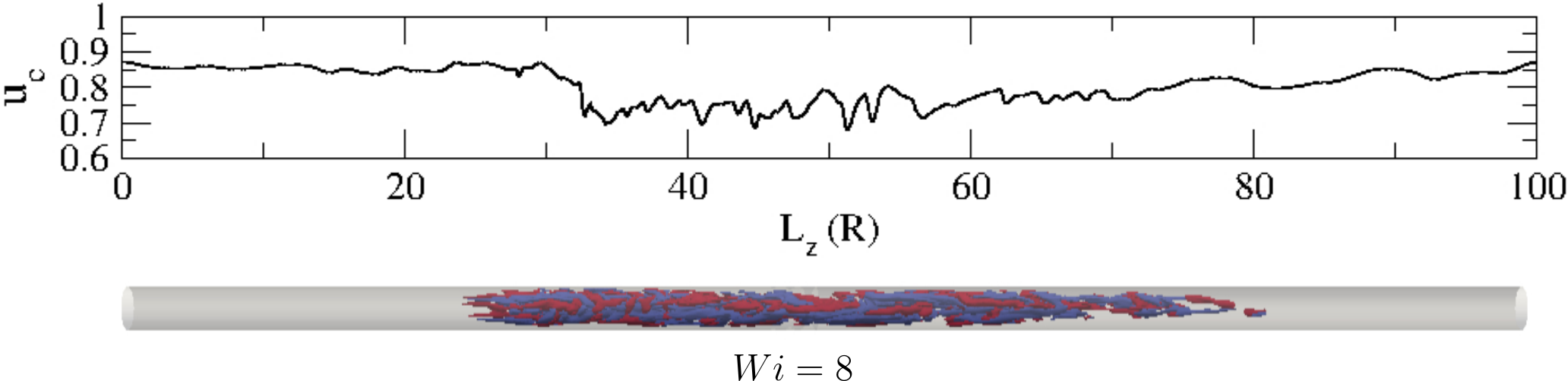} &
      \includegraphics[width=\piclen,height=3cm]{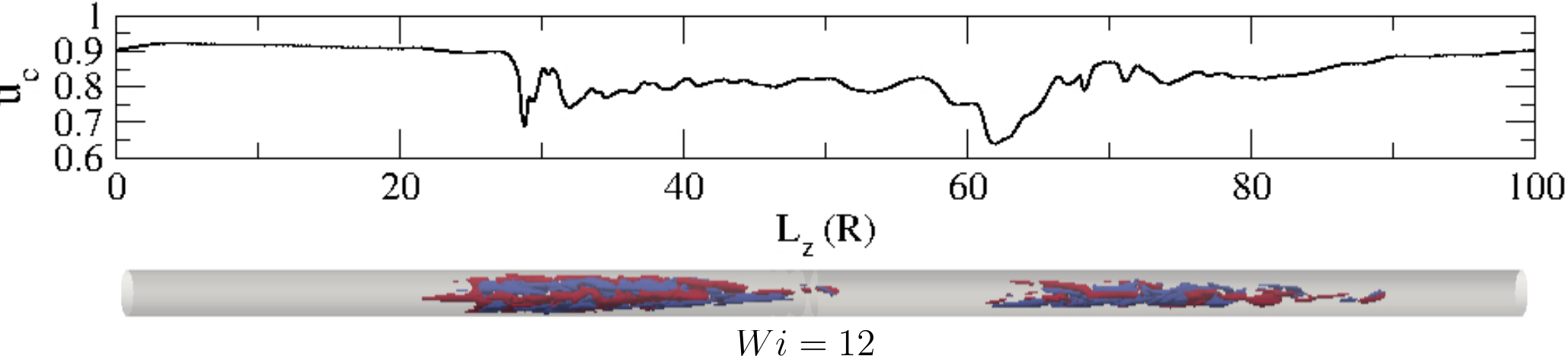}\\
      \multicolumn{2}{c}{$(e)$}\\
      \multicolumn{2}{c}{\includegraphics[width=\piclen,height=3cm]{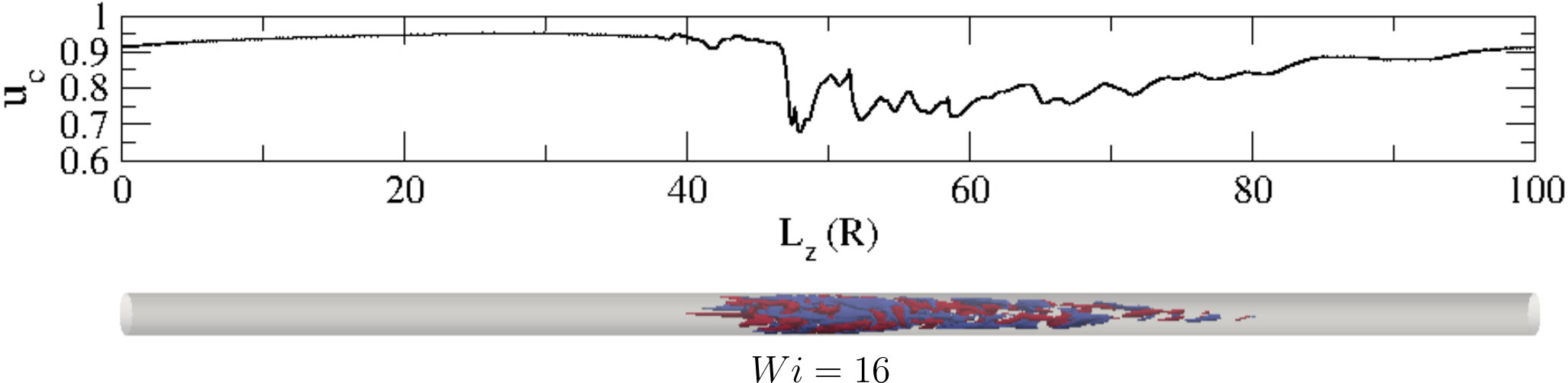}}\\
    \end{tabular}
  \end{center}
  \caption{ (Color online) Evolution of the spatio-temporal dynamics and turbulence structures
    as $\W$ increases when the simulations are carried out in a long pipe of $100R$ in
    the streamwise direction. The flow direction is from left to right.
    Note that the aspect ratio of the pipe has been increased to facilitate visualisation
    of the structures.}
  \label{fig:reverse}
\end{figure}

To assess the influence of the pipe length in the results of \S~\ref{sec:short},
the same computational procedure was repeated using larger pipes ($20R$ and $40R$).
A comparison of the drag reduction scenario obtained for the LE polymer when
the pipe length was varied is shown in figure~\ref{fig:drag_reduction_simu} $(b)$.
A first interesting observation is that, consistent with other works on viscoelastic
turbulence~\citep{Li06,Wang17}, viscoelasticity increases the streamwise correlation length with respect to
Newtonian simulations. Hence, while at these low Re a streamwise length of $10R$ is
enough to obtain realistic statistics in Newtonian pipe flow, viscoelastic simulations are still affected by streamwise periodicity
and result in lower drag reduction than those obtained
when larger pipes are used. Simulations performed in $20R$ and $40R$ long pipes produce
nearly the same drag reduction up to $\W \sim 6.75$, but differ both quantitatively and
qualitatively when relaminarisation is approached. For simulations using a $20R$ long pipe,
the same qualitative scenario as in the 10R long pipe simulations
is found: the drag reduction remains nearly constant over a finite range of $\W$, 
$6 \le \W \le 8$, before relaminarisation takes place. However, when a pipe of 40R is used,
this AR disappears and the drag reduction increases monotonically with
increasing $\W$ until the flow relaminarises.
This observation suggests that rather than being a manifestation of MDR,
the AR might be a consequence of the streamwise periodicity imposed in
the simulations and thus it might lack practical significance.

An additional test to confirm that the dynamics at the AR is different from that at
MDR is to compare the flow structures in our simulations
with recent experimental visualizations of MDR structures in pipe flow
at low Reynolds numbers~\citep{Cho18}. These experiments showed
that turbulence at MDR substantially differs from Newtonian type turbulence and it
is characterized by very elongated streaks which are  slightly inclined away from
the wall (see figure 3 in \cite{Cho18}). If the dynamics at the AR corresponded to MDR,
similar flow structures should be observed in our simulations, provided that the computational domain
is long enough to accomodate them. To examine this possibility, we have performed
a new set of simulations using a pipe of $100R$ in axial direction, which is approximately twice 
the size of the shortest structures observed by~\cite{Cho18}. Figure~\ref{fig:reverse}
illustrates the dynamical evolution of the turbulence structures as
$\W$ was increased in these simulations. It shows, at a certain
time instant, the variation of the centreline velocity $u_c$ along the pipe (top panel)
and isocontours of the radial velocity $w$ (bottom panel) for several $\W$ representative of
different dynamical regimes in the ME case. Note that a Newtonian case (fig.~\ref{fig:reverse} $(a)$)
has also been included for comparison. At low drag reduction ($\W < 6$), the dynamics
is very similar to that of the Newtonian case (see panels $(a)$ and $(b)$ in the figure).
Turbulence always fills the pipe entirely and the centreline velocity exhibits
comparable fluctuation levels in both cases. Nevertheless, the flow structures
in the viscoelastic case are broader and slightly more elongated in the axial direction
than those in pure Newtonian turbulence, reflecting the drag reduced nature of the
flow in viscoelastic simulations. Another clear distinction is that, while turbulence
extends across the entire pipe diameter in the Newtonian case, there are
several areas in the drag reduced flow where the near wall turbulence has been
suppressed by polymer activity. As $\W$ increases (between $\W= 7$
and $11$), the dynamics exhibit a complex spatio temporal behaviour.
As shown in fig.~\ref{fig:reverse} $(c)$ for $\W=8$, turbulence is confined to streamwise
localised patches known in the Newtonian pipe flow literature as \emph{slugs}. The
distance between the turbulent fronts, i.e. the interfaces separating laminar from turbulent flow,
increases progressively with time until the turbulence eventually fills the entire pipe. This 
space-filling turbulent state does not persist long and turbulence takes back the form
of \emph{slugs}, thereby restarting the cycle again. With further increase in $\W$,
coinciding with those $\W$ at which the AR occurs in shorter pipes, $12 \le \W \le 16$, turbulence becomes
permanently localised in the streamwise direction taking the form of turbulent \emph{puffs}.
As seen in fig.~\ref{fig:reverse} $(e)$, these viscoelastic \emph{puffs} are very similar to
Newtonian \emph{puffs}:  arrow-headed structures where turbulence is mainly concentrated in
the sharp upstream edge and progressively diffuses away as the puff is followed downstream. Unlike \emph{slugs},
\emph{puffs} keep their size constant and travel downstream at a nearly constant speed. We also found that
these \emph{puffs} sporadically split into two smaller \emph{puff}-like structures (see fig.~\ref{fig:reverse} $(d)$).
However, since the domain is not large enough to contain two full-size \emph{puffs}, there is a strong interaction between them
which causes the downstream \emph{puff} to quickly relaminarise~\citep{Hof10}. We note here that, although the pipe length
    in these simulations is enough to identify spatially localized structures, these are still affected by the finite size of the
    computational domain. As a result, laminar flow is not fully recover, i.e. the centreline velocity does not recover its laminar
    value $u_c = 1$, and the length of the simulated puffs is slightly shorter than that in laboratory experiments.  Finally, when $\W$ is increased above $16$
the flow fully relaminarises, showing that this is a robust feature of these simulations
which occurs at the same $\W$ regardless of the pipe length considered.

The dynamical scenario described above raises two important points. Firstly,
increasing $\W$ in these simulations leads to a relaminarisation scenario which follows
the same sequence of states as the transition to turbulence in
the Newtonian case but in reverse direction, i.e. turbulence,
slugs, puff splitting, puffs and laminar flow. We will henceforth
refer to the dynamics of this relaminarisation scenario as reverse
transitional dynamics. Note that in Newtonian pipe flow turbulence,
\emph{puffs} and \emph{slugs} are only found in the transitional regime
at significantly lower Reynolds numbers, $1800 \lessapprox \R_{puffs} \lessapprox 2300$ and
$2300 \lessapprox \R_{slugs} \lessapprox 2900$, than in these
viscoelastic simulations where $\R = 3500$. The effect of viscoelasticity
can thus be interpreted as a shift of the transition scenario of Newtonian pipe
flow turbulence towards larger $\R$. Secondly,
the dynamics at the $\W$ corresponding to the AR, $12 \le \W \le 16$,  is characterized by \emph{puffs}
and this is qualitatively very different from the structures observed at MDR in experiments.

\section{Comparison with experimental results}\label{sec:experimental}

\begin{figure}
  \begin{center}\setlength{\piclen}{\linewidth}
    \begin{tabular}{cc}
      \multicolumn{2}{c}{$(a)$}\\
      \multicolumn{2}{c}{\includegraphics[width=\piclen]{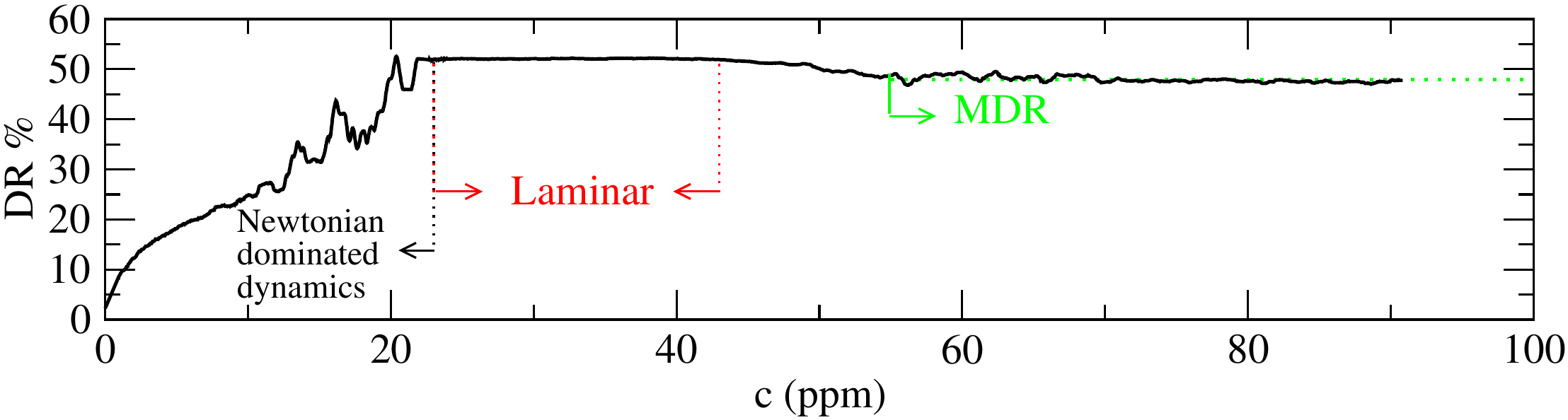}}\\
      $(b)$ 0 ppm, Newtonian & $(c)$ 15 ppm, slug\\
      \includegraphics[width=0.5\piclen]{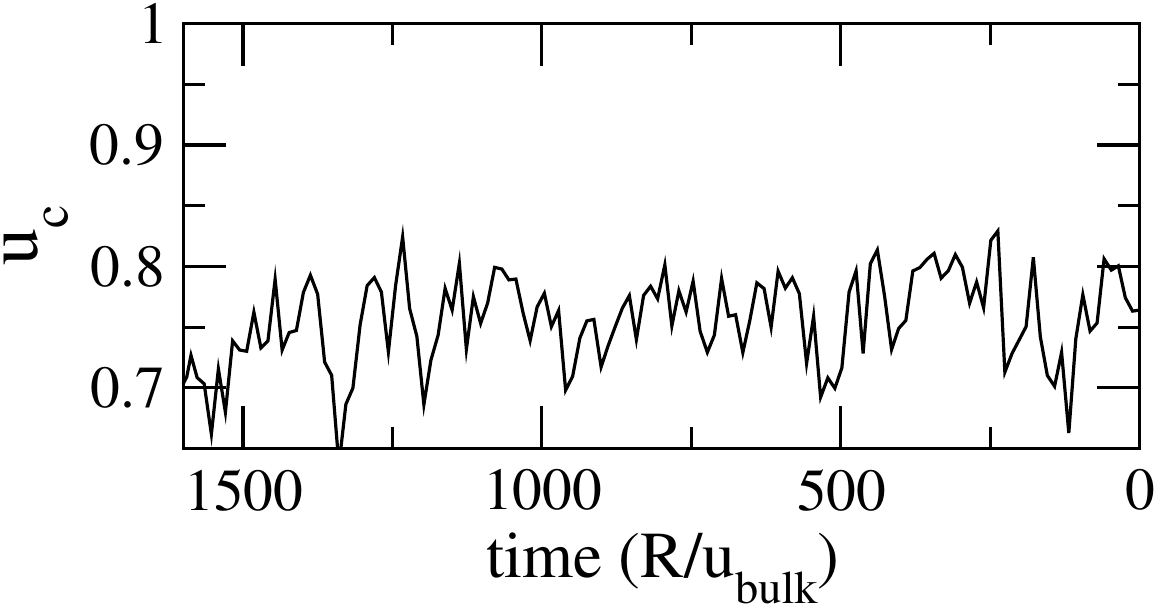}&
      \includegraphics[width=0.5\piclen]{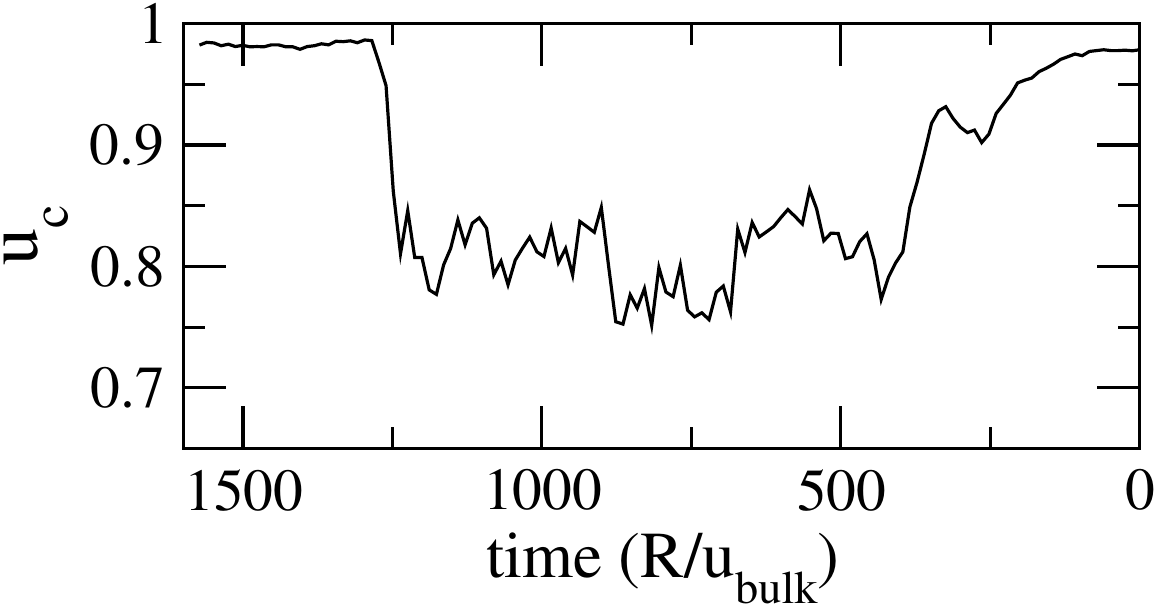}\\
      $(d)$ 20 ppm, puff & $(e)$ 20 ppm, splitting\\
      \includegraphics[width=0.5\piclen]{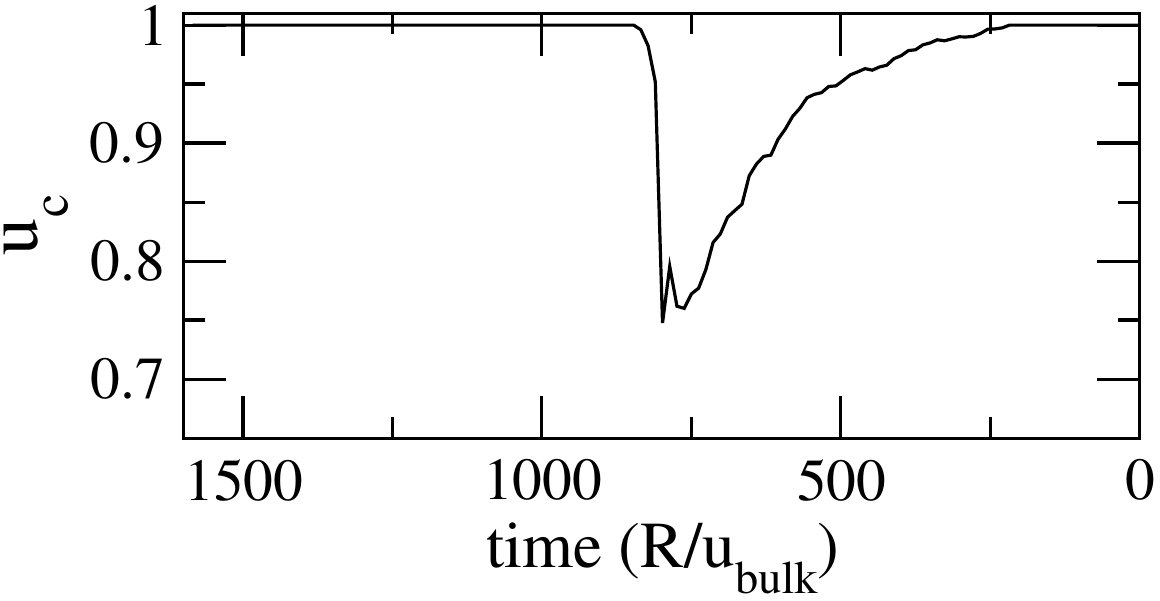}&
      \includegraphics[width=0.5\piclen]{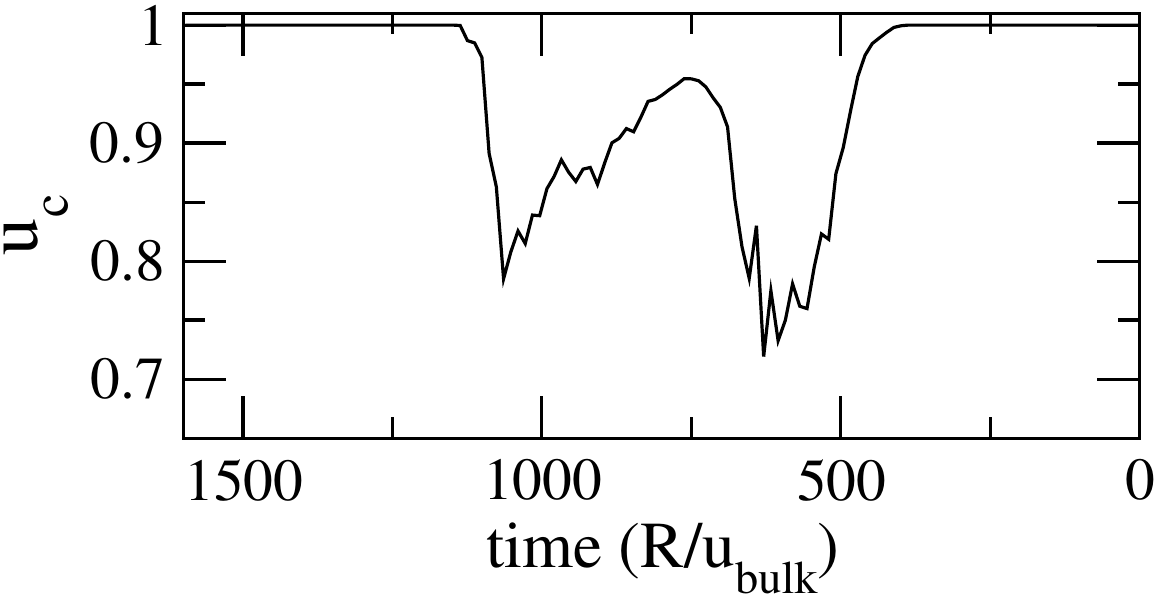}\\
      \multicolumn{2}{c}{$(f)$ 70 ppm, MDR}\\
      \multicolumn{2}{c}{\includegraphics[width=0.5\piclen]{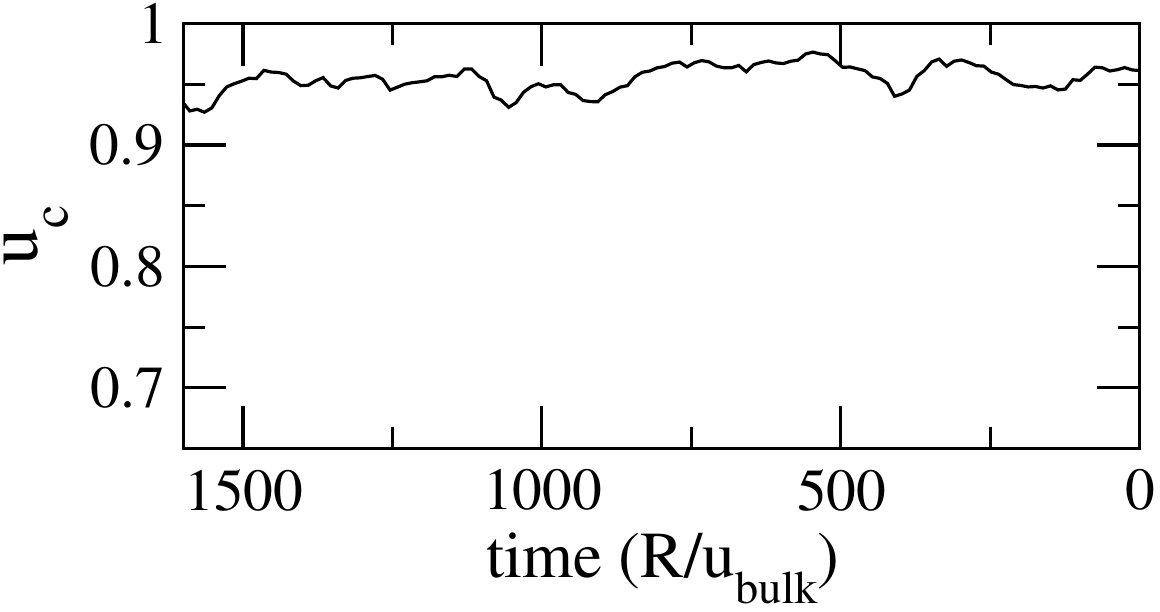}}\\  
    \end{tabular}
  \end{center}
  \caption{ $(a)$ Evolution of the drag reduction percentage DR\% with increasing polymer concentration (expressed in
    parts per million by weight ppm) in the experiments of C,L\&H for $\R = 3150$. $(b)-(f)$ LDV measurements
    of the centreline velocity $u_c$ illustrating the changes in the dynamics as polymer concentration increases.}
  \label{fig:experimental}
\end{figure}

The question now is whether the reverse transitional dynamics captured by our
simulations provides a meaningful description of viscoelastic pipe flow dynamics, i.e.
whether or not these simulations reproduce experimental observations. To answer this question
we provide in this section a detailed description of the dynamical scenario
found by \cite{Cho18} in pipe flow laboratory experiments at a similar
Reynolds number, $\R=3150$, when the  polymer concentration $c$ is increased progressively
from Newtonian turbulence to MDR (for details about the experimental setup, see
supplementary material in \cite{Cho18}). Note that the control parameter in these experiments
is polymer concentration, whereas in simulations it is the polymer relaxation time $\lambda$, i.e. $\W$,
that varies. These two magnitudes are however directly correlated. It has been shown that
even in dilute polymer solutions the relaxation time grows
with increasing polymer concentration~\citep{relaxation17}. Hence, increasing $\W$ in our simulations is
related to increasing polymer concentration in experiments.
Figure~\ref{fig:experimental} $(a)$ shows the variation
of the drag reduction percentage as polymer concentration was varied in the experiments.
Similarly to what occurs in the simulations, the amount of drag reduction $DR\%$
increases initially with increasing polymer concentration until a threshold value
is reached, $c \sim 23$ ppm (parts per million by weight), at which the flow fully relaminarises.
The flow remains laminar regardless of the imposed perturbations
over a significant range of polymer concentration ($c \sim 23-43$ ppm).
However, for  $c > 43$ ppm,  the flow becomes chaotic again and the
drag reduction level approaches progressively the Virk's asymptote.

Panels $(b)$ to $(f)$ in figure~\ref{fig:experimental}
illustrate how the dynamics change as the polymer concentration increases. More specifically,
these figures show the temporal variation of the centreline velocity $u_c$
obtained from LDV measurements at a central streamwise location.  The
x-axis has been inverted to facilitate comparison with the
instantaneous streamwise distribution of $u_c$ shown in figure~\ref{fig:reverse}. Note that,
as in the simulations,  $u_c$ is normalized with the centreline velocity of the
laminar state. In the absence of polymers
(see figure~\ref{fig:experimental} $(b)$)
the flow is fully turbulent and $u_c$ exhibits persistent random amplitude fluctuations.
As the polymer concentration is increased ($c \ge 13$ ppm), time intervals where $u_c$ strongly fluctuates
alternate with others at which it nearly recovers its laminar value
(see figure~\ref{fig:experimental} $(c)$  and compare to the analogous case in the
  simulations, figure~\ref{fig:reverse} $(c)$).
This temporal intermittency between turbulent and laminar states indicates that
the dynamics at this regime is characterized by spatially localized
structures. Furthermore, since the duration of these turbulent and laminar
intervals is highly variable, and both trailing and leading edge interfaces
show a sharp adjustment of the centreline velocity,
it is evident that these localised structures correspond to
\emph{slugs}~\citep{wygnanski_champagne_1973}.
With further increase in concentration ($c \ge 18$ ppm),
\emph{slugs} are replaced by \emph{puffs} (see figure~\ref{fig:experimental} $(d)$  and
analogous case in the simulations, figure~\ref{fig:reverse} $(e)$).
These structures are clearly distinguishable because of their long diffusive
tail and sharp velocity variation associated with the upstream edge.  As occurs
in the simulations  (figure~\ref{fig:reverse} $(d)$), splitting events are also frequently encountered
in the experiments (see figure~\ref{fig:experimental} $(e)$), leading either
to the emergence of \emph{slugs} or trains of puffs depending on the polymer
concentration. When $23$ ppm $ < c < 43$ ppm, turbulence is fully
supressed by the polymers and $u_c$ remains constant and equal to the laminar value.
It should  be emphasized at this point that the dynamics taking
place in the experiments is in excellent qualitative agreement with the reverse transition
found in the simulations. Ultimately, for $c \ge 43$ ppm
(see fig.~\ref{fig:experimental} $(f)$), the flow reaches MDR and $u_c$
exhibits again persistent oscillations. The frequency and amplitude of these oscillations
are however much lower than those for a fully turbulent Newtonian flow,
and the deviation of $u_c$ from laminar flow always remains less than $10\%$.     

The existence of a wide range of polymer concentrations at which the flow is laminar
makes a clear distinction between two regimes where polymers play
different dynamical roles. In the  first regime the role of the polymers is
to suppress turbulence and cause a reverse transition. As discussed in \S\ref{sec:large},
the dynamics in this regime are dominated by the same flow
structures as in the Newtonian case and polymers simply act
to delay the transition scenario. In the second regime, for $c > 43$ ppm,
the interplay between high polymer elasticity and inertial effects
drives an instability, dubbed in Samanta et al. (2012) as
elasto-inertial instability (EII), which results in a new turbulence type,
elasto-inertial turbulence (EIT). As shown in fig.~\ref{fig:experimental} $(a)$,
the drag reduction level associated with EIT closely matches that of the
Virk's asymptote and  it remains unchanged as polymer concentration increases.
These observations strongly suggest a direct link between EIT and
MDR, thereby offering an explanation to the universality of
this asymptotic limit. An additional remark about EIT (and thus MDR)
is that as seen in figure~\ref{fig:experimental} $(f)$, it is always space-filling
and no spatio-temporal intermittency is observed in this regime.
This is an important feature that can help distinguish realistic
MDR dynamics from other regimes with similar statistical properties.
An example of the latter are the puffs found prior to
relaminarisation. We found in both simulations and experiments
that the average friction coefficient and mean velocity profiles associated
with these puffs are nearly identical to those at MDR (not shown). Hence, the circumstance
that time averaged statitistical quantities match those of MDR
(main criterion to identify MDR in many earlier studies) is a necessary but not sufficient condition
to identify this regime in numerical simulations. An analysis of the spatio-temporal dynamics
must also be carried out to discern whether or not the simulated
flows belong to the MDR regime.

\section{Elasto-inertial turbulence}

\begin{figure}\setlength{\piclen}{0.6\linewidth}
  \begin{center}
    \begin{tabular}{cc}
      $(a)$ & $(b)$\\
      \includegraphics[width=0.6\piclen]{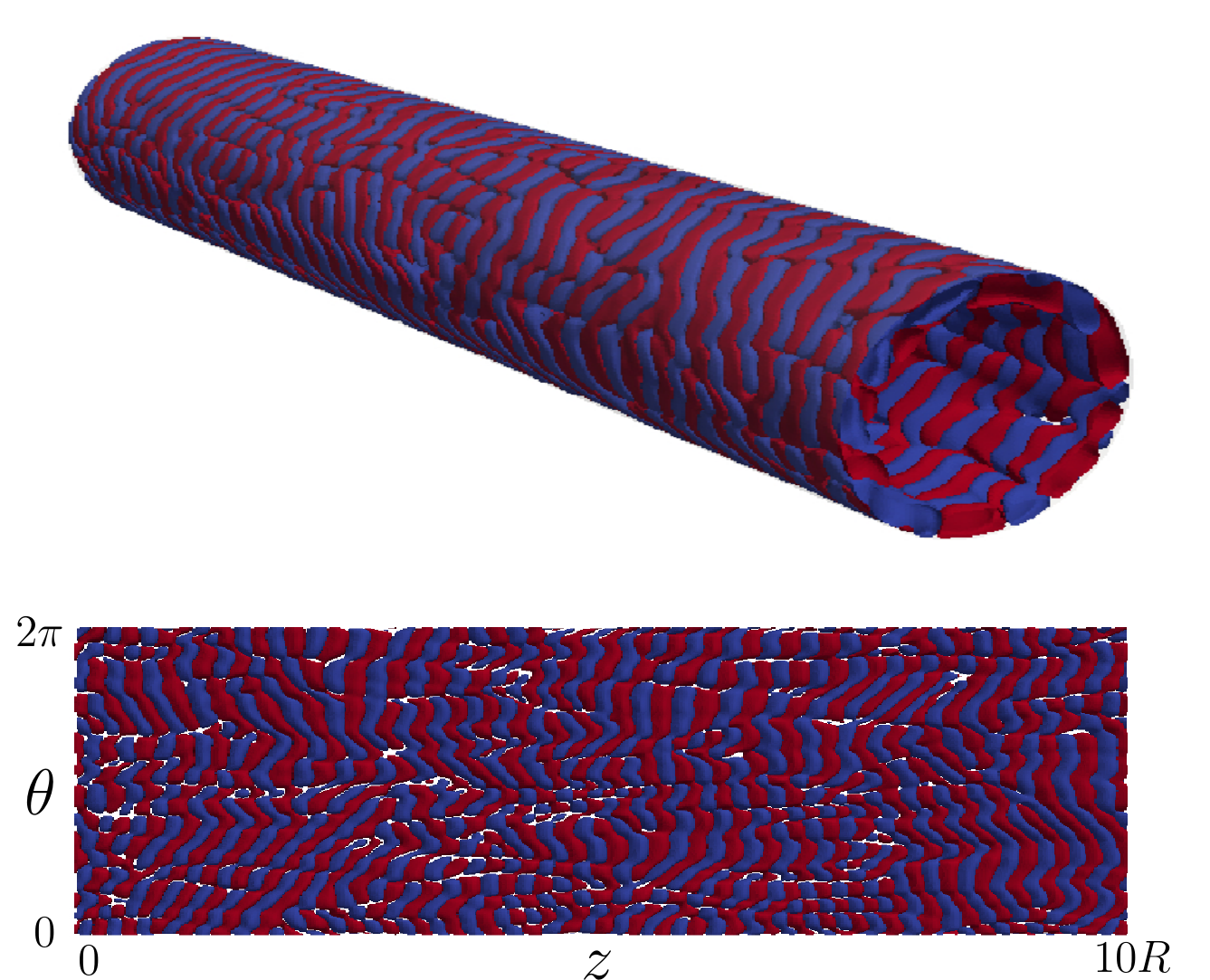} &
      \includegraphics[width=\piclen]{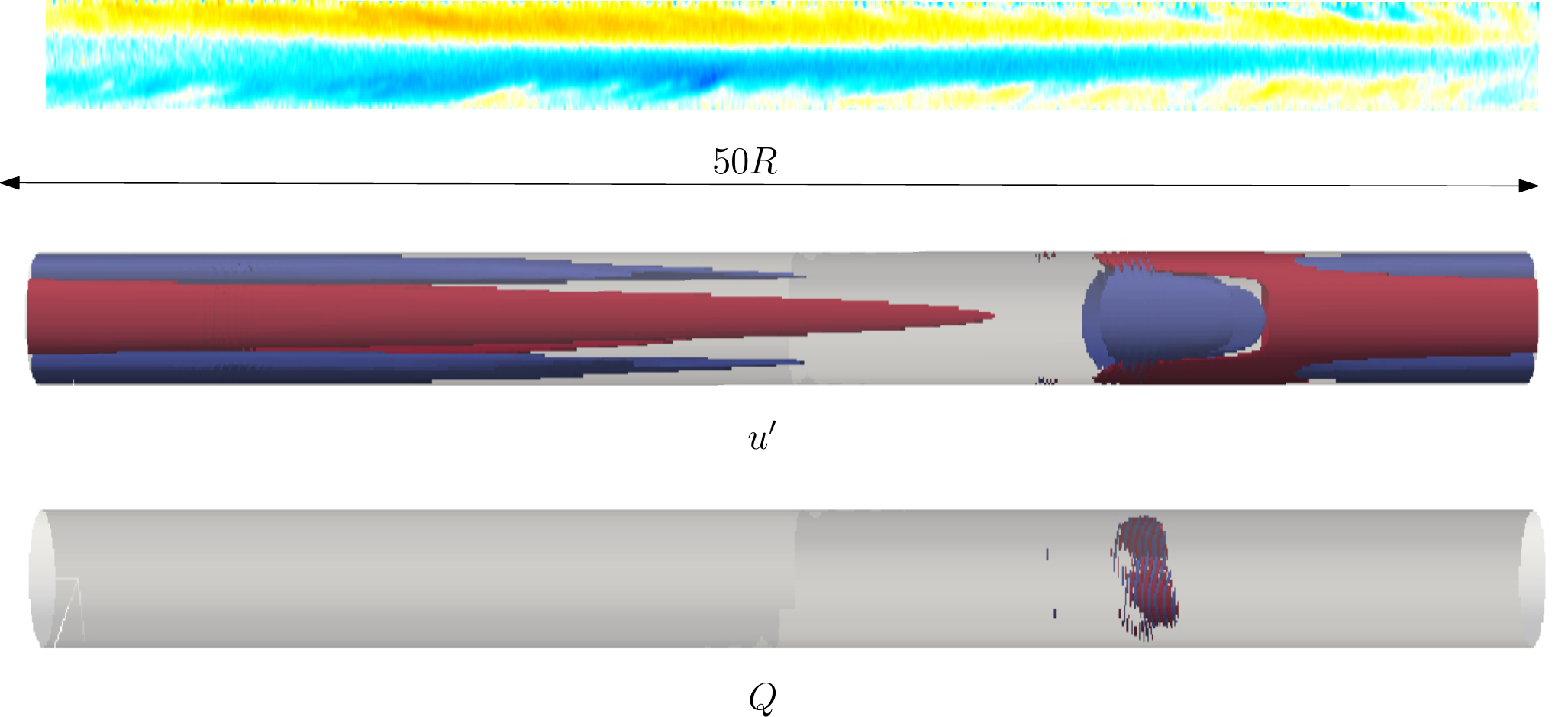}\\
    \end{tabular}
  \end{center}
  \caption{ (Color online) $(a)$ Isosurfaces of the second-moment of the velocity gradient tensor $Q=-0.005$ and
    $Q= 0.005$ illustrating the topological structure of elasto-inertial turbulence. The top panel
    shows a 3D view of the pipe highlighting the near wall localisation of the structures. The bottom panel shows the
    characteristic pattern,  with alternating regions of rotational and extensional/compressional behaviour,
    in a cylindrical section $\theta-z$ at the wall. The state shown
    corresponds to a simulation conducted at $\W=60$ for the LE case. The second invariant of the
    velocity gradient tensor is computed as $Q = (1/2) (||\Omega||^2 - ||\Gamma||^2)$, where $\Omega = (1/2) (\nabla \mathbf{v} - \nabla \mathbf{v}^T) $ is the
    vorticity tensor and $\Gamma = (1/2) (\nabla \mathbf{v} + \nabla \mathbf{v}^T) $ is the rate-of-strain tensor. $(b)$ The top panel shows the deviation
    of the streamwise velocity from the mean flow $u'$ in a state of EIT for experiments conducted at $\R=3150$. The velocity
    was measured using PIV in a pipe cross-section of nearly $6R$ in the axial direction. The image
    shown was obtained by assuming the Taylor's frozen hypothesis, i.e. turbulence is advected downstream quickly
    and changes in time are slow. The intermediate and bottom panels show $u'$ and $Q$ respectively
    for a simulation performed at $\R=3500$ and $\W=30$ in a $40R$ long pipe. Two isocontours $u'=\pm0.1u'_{max}$ and
    $Q=\pm0.005$ were used in each case.}\label{fig:elasto}
\end{figure} 

We have shown so far that our FENEP-NS simulations
qualitatively reproduce the dynamics observed in experiments
up to the point where relaminarisation occurs. The next question is therefore
whether by increasing $\W$ beyond the relaminarisation threshold
these simulations are  also capable of capturing the EII and MDR.
To address this question we have performed several simulations at
$\W$ ranging from $20$ to $80$ in both the ME and LE cases.
The pipe length was initially set again to $10R$. The simulations were initialized from the base flow,
previously computed, which was perturbed by adding a pair of streamwise localized
rolls ($v=A(g+rg') cos(\theta) e^{-10sin^2(\pi z/L_z)}$ and
$w=A g sin(\theta) e^{-10sin^2(\pi z/L_z)}$, where $g=(1-r^2)^2$ and $A$ is the amplitude of the
disturbance). In all simulations carried out for the ME case,
the energy of the disturbance grows initially due to the lift-up mechanism,
but after approximately 150 $R/u_{lc}$ time units it decays gradually with time and
the flow fully relaminarises. For the LE case, however, we find that a secondary instability
sets in for $\W \geq 30$. While similarly to the ME case transient growth and
subsequent decay in energy are initially observed, here the energy increases again
as the time evolves and eventually saturates to a new flow state significantly less energetic
than that of Newtonian type turbulence. A possible explanation for this
behaviour is as follows. Due to the initial disturbance polymers are
greatly stretched and accumulate a significant amount of elastic energy.
In response to this stretch, polymers generate stresses which act to weaken and eventually
suppress this turbulence. As the turbulence intensity decays,
polymers relax and the elastic energy they store is progressively transferred to the fluid.
As a result, the kinetic energy increases again and a new form of instability takes place.
The topological structure of the new flow state is illustrated in figure~\ref{fig:elasto} $(a)$
through isocountours of the second invariant of the velocity gradient tensor $Q$.
Note that this quantity has been chosen to facilitate comparison with other works
on EIT~\citep{Samanta13,dubief13}.
Regions of intense vorticity ($Q > 0$, red) are found to
alternate with strain-dominated regions ($Q < 0$, blue) creating a chaotic pattern of
elongated spanwise oriented structures aligned in streamwise direction.
The vortices are localized in the near wall region and are essentially two-dimensional with rotation being in the $r-z$ plane.
We note that this spatial arrangement of structures in the near wall region is  very different
from Newtonian type turbulence, where the dominant structures are oriented in the streamwise direction.
This flow state reproduces two  essential features of MDR: the drag reduction level associated with this state
remains nearly constant as $\W$ increases, and although the average friction factor ($f \sim 0.0047$)
is slightly below that corresponding to the Virk's asymptote ($f = 0.0051$), it is reasonably close to it. It
should be noted that the Virk's asymptote is a fit of empirical data collected from different
experiments. As such, it should be used as an estimate for the friction of the MDR state
rather than as a categorical result. All these observations are consistent with previous reports
of elasto-inertial turbulence in channel flow simulations~\citep{Samanta13,dubief13}.

The top panel in figure~\ref{fig:elasto} $(b)$ illustrates the typical flow structures of EIT in
the experiments. It shows the streamwise velocity deviation with respect to the
mean flow $u'$ over a length of $50R$. Note that in the top panel the velocity was obtained
from particle image velocimetry (PIV) in a section of nearly $6R$ in the axial direction
and the Taylor's Frozen turbulence hypothesis was then assumed to reconstruct the structures
shown. As seen, the structure of EIT is clearly dominated by very elongated streaky
structures aligned in the flow direction with a slight slope towards the centreline.
The axial length of these structures is highly variable,
ranging approximately from $50R$ to $200R$, being more elongated near the instability onset.
As polymer concentration increases, the structures become shorter and increasingly more
chaotic but still preserve their characteristic inclination.
Unlike in the simulations, vortical structures could not be resolved in the near wall
region in the experiments. The vortical structures observed in the simulations are considerably
weaker than Newtonian flow structures, which makes a detection in experiments difficult.
In addition they are located close to the wall where the measurement accuracy is lower. In the simulations the problem
is the opposite. Because of the Gauss-Lobatto-Chebyshev grid used in the radial direction
the computational nodes are clustered near the wall, enabling an accurate resolution of the flow in this area. 
Nevertheless, the necessity of very dense grids in the streamwise direction to properly
resolve the near wall structures makes it extremely costly to use axial domains
sufficiently large as to capture the large scale structures observed in the experiments.
A direct comparison of the structure of EIT between experiments and simulations is thus challenging.
It is however tempting to investigate whether large scale structures can also be identified in simulations,
and if their length approaches that of the structures in experiments as the computational domain is
increased. To that extent, we have performed an additional simulation at $\W=30$ using
a pipe of $40R$ in streamwise direction. EIT could only be captured transiently
in this simulation and after approximately $2500$ time units the flow went back to laminar.
%We believe that the transient nature of the instability in this larger domain
%is because,  despite having used a fine grid in streamwise direction,
%the spatial resolution is still insufficient to properly capture the near wall
%structures.
Nevertheless, some interesting dynamical aspects could be inferred from
this simulation. As seen in the bottom panel of figure~\ref{fig:elasto} $(b)$,
if the same threshold $Q=\pm 0.005$ as in figure figure~\ref{fig:elasto} $(a)$ is used,
the near wall vortices appear localized over a short region of nearly $2R$
in the streamwise direction. Large scale streamwise velocity structures (see intermediate panel)
seem to emerge from the area where the vortices are located and extend almost over
the entire domain. These structures become thinner as they are followed
downstream and take an arrow shape at the leading edge which closely
resembles the inclination away from the wall observed in the experiments. 
This structural similarity between EIT in simulations and experiments
suggests that the flow in both cases may be driven by the same
instability. However, the precise dynamical relation between
the small near wall structures and these elongated streaks still
remains to be determined and will be the focus of a future
investigation.

\section{Conclusions}

%The purpose of this work was to clarify the link between
%relevant numerical findings on viscoelastic channel flow
%simulations, such as hibernating turbulence or elasto-inertial
%turbulence, with recent experimental observations in
%viscoelastic pipe flow. To that extent, we have performed
%direct numerical simulations ofviscoelastic pipe flow using
%the FENEP model for polymer dynamics 
%Direct numerical simulations of viscoelastic pipe flow using the FENEP
%model for polymer dynamics have been performed at $\R=3500$  

We have investigated numerically the dynamics of viscoelastic
pipe flow at $\R=3500$, where in the Newtonian case
flows are fully turbulent~\citep{BaSonLeAvHo15}. In agreement with recent experimental
observations, we find that the dynamics as $\W$ increases can be categorized
in two regimes. The first regime takes place for low-to-moderate
$\W$ and the dynamics are essentially of the Newtonian type. The influence
of polymers on this regime manifests itself as a shift of
the transitional scenario towards larger Reynolds
numbers. As a result, as $\W$ increases, the flow transitions
from turbulence to laminar following the same stages as in
the Newtonian turbulence transition, but in reverse order,
i.e. fully turbulent, slugs, puff splitting, puffs and laminar.
The second regime occurs at large $\W$ and could
only be captured in the simulations when considering polymers with
very large extensibility. The amount of drag reduction
associated with this regime nearly matches
that of the Virk's asymptote and remains unchanged as $\W$ increases.
This strongly suggests a direct link between this regime
and MDR. Separating these two regimes there is a
significant range of $\W$ for which the flow relaminarises regardless of the
initial condition. The existence of this laminar regime implies that the dynamics
at the elasticity dominated regime is disconnected from Newtonian type
turbulence, and consequently it would have to originate from a separate instability (EII).
While experiments cannot resolve the small vortical structures characteristic for EIT in simulations,
the large scale inclined streaks seen in experiments are also present in the simulations. 
It remains for future investigations to establish the link of these streaks with the near wall vortices.\\

\noindent We also show that MDR in simulations cannot be identified based on average profiles and
friction values alone. While in the hibernating regime these quantities are close to those of MDR,
larger domain studies identify this regime as spatio temporal intermittency and as part of a reverse transition scenario.
The asymptotic MDR regime is only approached for even larger Weissenberg numbers. 

%\noindent We also show that the streamwise length of the simulation domain plays
%an essential role in understanding the dynamics associated to viscoelastic
%simulations. In short domains, the reverse transitional dynamics
%manifests itself in the form of temporal intermittency, where periods of low friction
%or hibernation become more frequent and durable
%as $\W$ increases. Due to the statistical resemblance between these hibernation 
%periods and MDR, it has been speculated that MDR might simply be a marginal
%state of hibernating turbulence~\citep{XiGra10a}.
%Our simulations, however, reveal that these hibernating events cannot be related
%to MDR as they occur in the inertia dominated regime before relaminarisation takes place. 
%The increasing dominance of hibernating events with increasing $\W$ can
%be interpreted as a manifestation of spatial localization in simulations
%where the domain size is not large enough to capture spatially-localized structures.
%In other words, these hibernating events reflect the progressive increase in the
%fraction of laminar flow occurring as the flow transitions from space-filling turbulence
%to spatially-localized turbulence. Keeping all parameters constant and increasing
%the domain size result in puffs interspersed by quasi-laminar regions and identifies
%this $\W$ regime as part of the reverse transition that takes place with increasing
%$\W$.  

\bibliography{local}
\bibliographystyle{jfm}

\end{document}